\documentclass[12pt,preprint]{emulateapj}
\usepackage{graphicx}

\shorttitle{Decoupling Phase Variations}
\shortauthors{Stephen R. Kane \& Dawn M. Gelino}
\slugcomment{Submitted for publication in the Astrophysical Journal}

\begin{document}

\title{Decoupling Phase Variations in Multi-Planet Systems}
\author{Stephen R. Kane, Dawn M. Gelino}
\affil{NASA Exoplanet Science Institute, Caltech, MS 100-22, 770
  South Wilson Avenue, Pasadena, CA 91125, USA}
\email{skane@ipac.caltech.edu}


\begin{abstract}

Due to the exquisite photometric precision, transiting exoplanet
discoveries from the Kepler mission are enabling several new
techniques of confirmation and characterization. One of these newly
accessible techniques analyzes the phase variations of planets as they
orbit their stars. The predicted phase variation for multi-planet
systems can become rapidly complicated and depends upon the period,
radius, and albedo distributions for planets in the system. Here we
describe the confusion which may occur due to short-period terrestrial
planets and/or non-transiting planets in a system, which can add
high-frequency correlated noise or low-frequency trends to the data
stream. We describe these sources of ambiguity with several examples,
including that of our Solar System. We further show how decoupling of
these signals may be achieved with application to the Kepler-20 and
Kepler-33 multi-planet systems.

\end{abstract}



\section{Introduction}
\label{intro}

The past few decades represent an extraordinary period of growth in
our knowledge of planetary systems. Whereas previously we used the
single data point of our own Solar System to derive theories of
planetary formation, we now have a diverse range of planetary systems
to draw upon. We are gaining an improved understanding of the
distribution of planetary properties such as period, mass,
eccentricity, radius, and multiplicity. These last two have been aided
in no small part by the significant discoveries of the Kepler mission
which has revealed numerous cases of transiting planets in
multi-planet systems \citep{bor11a,bor11b,bat12}. Many of these are
tightly-packed systems of multiple planets within a relatively small
period range, examples of which include Kepler-11 \citep{lis11a},
Kepler-20 \citep{fre12,gau12}, and Kepler-33
\citep{lis12}. \citet{lis11b} and \citet{lis12} point out that these
multi-planet systems comprise a large fraction of the total candidates
detected and are far less likely to be due to false-positives.

The precision of the Kepler photometry has also allowed the
investigation of out-of-transit variations which are phased with the
planetary orbit. These include ellipsoidal variations of the host star
induced by the planet \citep{jac12,pfa08}, Doppler boosting or beaming
\citep{maz12,shp11}, and reflected light from the planet
\citep{kan10,kan11a,kan12}. A combination of these effects have been
detected in several cases, such as HAT-P-7b which was observed in
Kepler data to have signatures of reflected light as well as
ellipsoidal variations \citep{wel10}. These effects have almost
exclusively been considered for single-planet systems, usually in very
short period orbits where these effects will be higher in amplitude.

Here we consider the confusion that can result from multi-planet
systems when attempting to measure the photometric phase variations
resulting from reflected light. Since there will be a variety of
periods, radii, and albedos, the contribution of each planet to the
total phase variations may encompass a large range of amplitudes
depending on how these values are distributed. The result of this is
that certain planetary configurations will have ``rogue'' phase
variation components which masquerade as correlated noise in the
data. We describe a decoupling procedure for these systems to
disentangle the signatures of the individual planets. We discuss the
implications for the Solar System as an externally observed exosystem
and apply the method to the Kepler multi-planet systems of Kepler-20
and Kepler-33.


\section{Photometric Phase Variations}

The phase variations of exoplanets are described in detail from
numerous sources, such as those mentioned in Section \ref{intro}. We
refer the reader to \citet{kan10} and \citet{kan11a} since we will be
adopting that particular formalism.

\begin{figure*}
  \begin{center}
    \includegraphics[angle=270,width=15.0cm]{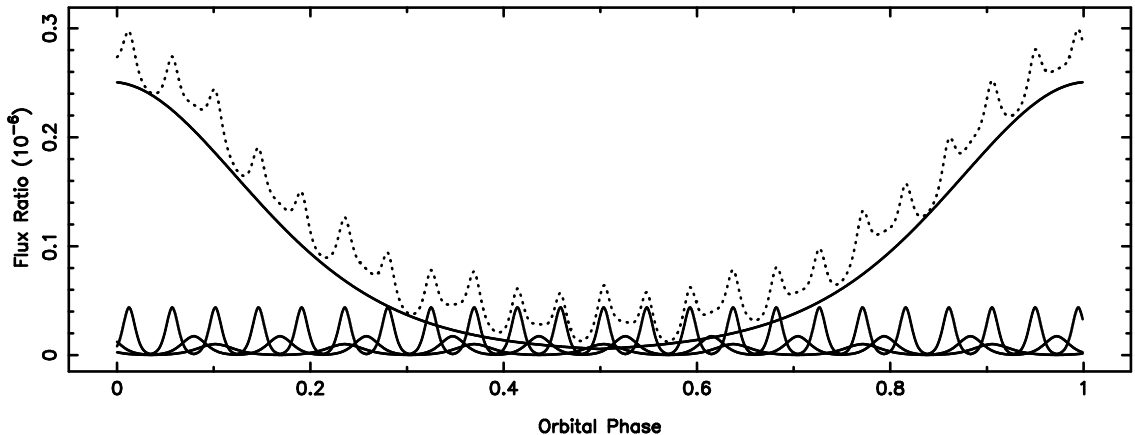}
  \end{center}
  \caption{The model photometric flux variations due to planetary
    phases in a hypothetical system consisting of three Earth-size
    planets in tightly-packed orbits close to the star and a
    Jupiter-size planet at 0.7~AU. The solid lines show the phase
    variations due to the individual planets and the dotted line
    indicates the combined effect.}
  \label{combo}
\end{figure*}

The flux ratio of an exoplanet to the host star as observed from Earth
is given by the following expression
\begin{equation}
  \epsilon(\alpha,\lambda) \equiv
  \frac{f_p(\alpha,\lambda)}{f_\star(\lambda)}
  = A_g(\lambda) g(\alpha,\lambda) \frac{R_p^2}{r^2}
  \label{fluxratio}
\end{equation}
where $\lambda$ is the wavelength of the observations, $A_g(\lambda)$
is the geometric albedo, $g(\alpha,\lambda)$ is the phase function,
and $R_p$ is the radius of the planet. The phase angle of the planet,
$\alpha$, is defined to be zero when the planet is at superior
conjunction. Here we also allow for orbital eccentricity by using the
time-dependent star--planet separation, $r$, given by
\begin{equation}
  r = \frac{a (1 - e^2)}{1 + e \cos f}
  \label{separation}
\end{equation}
where $a$ is the semi-major axis and $e$ is the orbital
eccentricity. Notice that Equation \ref{fluxratio} contains three
major components: the geometric albedo, the phase function, and the
inverse-square relation to the star--planet separation. For a given
phase function, the flux ratio is proportional to the albedo, the
square of the planetary radius, and the inverse square of the
separation. This is equivalent to being approximately proportional to
$P^{3/4}$ where $P$ is the orbital period. In other words,
\begin{equation}
  \epsilon(\alpha,\lambda) \propto A_g \times R_p^2 \times P^{3/4}
\end{equation}
The radius is clearly the dominant component which influences the
amplitude of the phase variation. As we will see, this is an important
factor when assessing the signatures of multi-planet systems,
particularly the tightly-packed systems detected by the Kepler
mission.


\section{Multi-Planet Systems}

Here we describe the combination effects and decoupling issues for
multi-planet systems.


\subsection{Combination Effects}

The discoveries from both RV and transit exoplanet detections have
revealed that planetary systems can come in a variety of orbital
configurations and physical characteristics. Multiplicity appears to
be a common trait of exosystems and thus we can expect most systems to
exhibit phase signatures of more than one planet. The manner in which
these signatures contribute to the total flux variations can cause an
ambiguous interpretation of the data. Short-period planets can cause
significant confusion if the observational cadence is insufficient to
sample the variation cycle caused by the planets. For systems in which
there is a dominant giant planet, the signatures of these smaller
short-period planets can appear as correlated noise in the data
depending on the photometric precision. In addition, it is unlikely
that all the planets in a given system will transit their host star
from the perspective of Earth, as we will see for the Solar System in
Section \ref{sstran}. Thus there may be additional unaccounted for
signatures in the data if one is basing the model on those known to
transit.

Shown in Figure \ref{combo} is an example model of flux variations due
to the phase signatures of a four-planet system, the contribution from
each planet shown as a solid line. The dominant source of the
reflected light is due to a Jupiter-size planet in a circular orbit at
0.7~AU. There are also three Earth-size planets with periods less than
30 days. The combined signature is represented by a dotted
line. Although one may detect the signal of the giant planet, the
terrestrial planets will add correlated noise to the signal, the
amplitude of which will depend on their respective albedos. The
opposite effect is also true, that the phase signature of three known
(from their transits) terrestrial planets will soon be distorted by
the presence of an unknown (non-transiting) exterior giant planet.


\subsection{Fourier Decoupling}
\label{fourier}

For multi-planet systems, the planets are more likely to be in
near-circular orbits than planets in single-planet systems
\citep{wri09}. This circularization of the orbits is even more common
for the tightly-packed planetary orbits found for the Kepler systems
\citep{mor11}. Thus, the combined phase signature of multi-planet
systems may be adequately described as a sum of trigonometric
functions suitable for fourier analysis. Consideration of phase
signatures for eccentric planets, such as those described by
\citet{kan10}, may also be considered in this context using Keplerian
techniques in the fourier method. This is particularly important when
evaluating the significance of peaks in the power spectrum and is
discussed in detail by \citet{cum04}, \citet{cum08}, and
\citet{oto09}.

\begin{figure*}
  \begin{center}
    \includegraphics[angle=270,width=15.0cm]{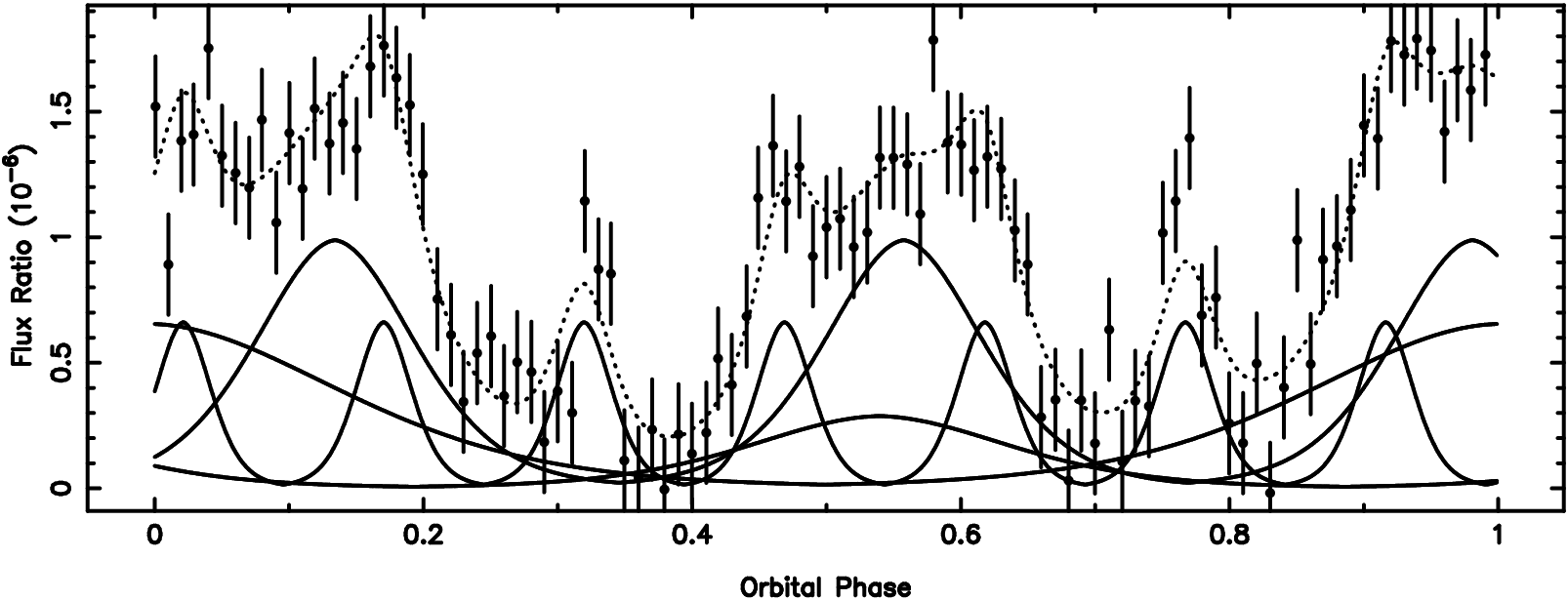} \\
    \includegraphics[angle=270,width=15.0cm]{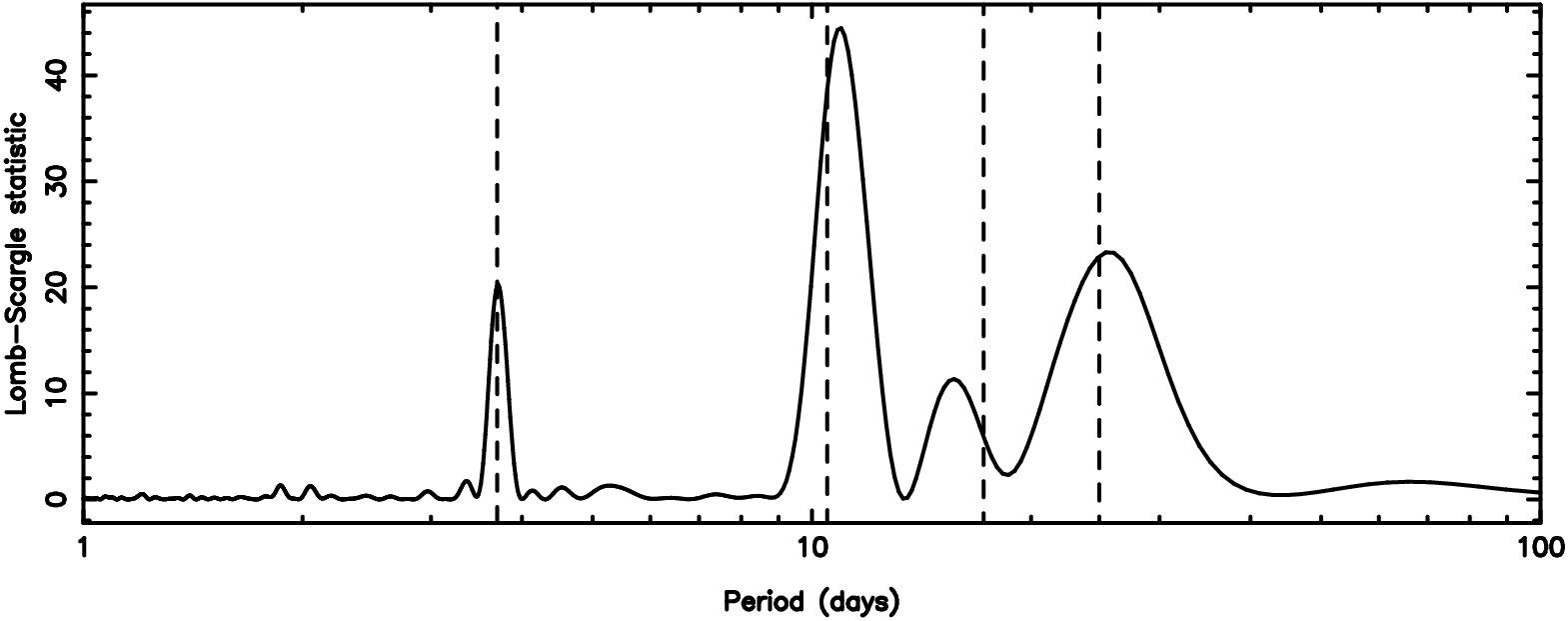}
  \end{center}
  \caption{Top panel: The flux variations exhibited by a hypothetical
    closely-packed system of four super-Earths. The combined signature
    of all four planets is shown as a dotted line, along with
    simulated data acquired over the time-span of one complete orbit
    of the outer planet. Bottom panel: The periodogram resulting from
    a fourier analysis of the data. The vertical dashed lines indicate
    the actual periods of the four planets.}
  \label{fourierfig}
\end{figure*}

The capability of a robust fourier analysis to recognize the
individual component planetary contributions to the total phase
variations depends primarily on (a) the photometric precision, (b) the
observational cadence, and (c) the duration of the observations. The
observational cadence and duration limits the period analysis between
the low fundamental frequency and the Nyquist frequency. A low
signal-to-noise dataset will produce peaks in the fourier spectrum of
similar power to spurious signals (aliases). An example is shown in
Figure \ref{fourierfig} where we have simulated a four-planet system
with radii ranging from super-Earth ($2R_\oplus$) to large ice giants
($5R_\oplus$). The orbits are tightly packed with the outer planet in
a $\sim 25$~day period orbit. The dotted line shows the combined phase
variation for the entire system of planets. We have simulated data for
200 epochs spread over twice the period of the outer planet. Each time
was passed through a gaussian filter with a standard deviation of 15
minutes to randomize the times. A weighted Lomb-Scargle (L-S) period
analysis \citep{lom76,sca82} was applied using the methodology shown
in \citet{aha05}. This method applies inverse variance weighting
calculated for each data point to each of the terms in the fourier
component. In cases where the error bars (and hence the weights) are
equal, the weighted L-S periodogram reduces to the standard L-S
periodogram. The bottom panel shows the periodogram for the simulated
data with vertical dashed lines indicating the actual periods of the
four-planet system.

There are a couple of issues to note regarding the Fourier
disentanglement of the individual signatures depicted here. The planet
located at $\sim 17$~days has the lowest phase amplitude, comparable
to the noise properties of the data, and therefore is barely detectable. As
such, improving either the cadence or the time baseline of the
observations has little effect in improving the detection of this
signature. The detection of the planet with the smallest period is
only affected in so far as the cadence drops to a level that the
orbital frequency becomes comparable to the Nyquist frequency. In this
case, that situation would result if the number of measurements
dropped from 200 to 27 yielding a cadence of $\sim 1.8$~days. One must
consider this detection threshold when planning the observation
strategy for a particular mission.

The fourier method is thus suitable to extract such signatures in most
cases except where the signal-to-noise is very low. Where one or more
phase signatures dominate the signal, an iterative approach of
subtracting those fitted signals and then re-analysing the residuals
can reveal the remaining planets. This method however does assume
gaussian noise with purely periodic signatures. Since the number of
cycles is very limited, the Maximum Entropy Method (MEM) is an
alternative approach since this is relatively efficient in detecting
frequency lines with few assumptions regarding the initial estimates
of the fit parameters (see for example \citet{end00}).


\section{The Solar System}

To find a classical case where phase variation confusion can occur, we
need look no further than our own Solar System. The data for the Solar
System planets were extracted from the JPL HORIZONS
System\footnote{\tt http://ssd.jpl.nasa.gov/?horizons}.


\subsection{Combined Phase Variations}

First we consider the total phase variations expected by the Solar
System when viewed along the ecliptic. Table \ref{sstable} contains
the Solar System orbital parameters used for these calculations. Note
that when treating the Solar System as an exosystem, one must be
careful to distinguish between the longitude of perihelion and the
argument of perihelion since the longitude of perihelion is the sum of
the longitude of the ascending node and the argument of
perihelion. Here we use the argument of perihelion ($\omega$) derived
from these quantities. From these parameters and the geometric
albedos, we calculate the predicted phase variations whose amplitudes
are also shown in Table \ref{sstable}.

\begin{deluxetable*}{ccccccccc}
  \tablecolumns{9}
  \tablewidth{0pc}
  \tablecaption{\label{sstable} Solar System Planetary Orbital
    Parameters and Peak Flux Ratios}
  \tablehead{
    \colhead{Planet} &
    \colhead{$P$} &
    \colhead{$a$} &
    \colhead{$e$} &
    \colhead{$\omega$} &
    \colhead{$R_p$} &
    \colhead{$i$} &
    \colhead{$A_g$} &
    \colhead{Flux Ratio} \\
    \colhead{} &
    \colhead{(days)} &
    \colhead{(AU)} &
    \colhead{} &
    \colhead{(deg)} &
    \colhead{($R_\oplus$)} &
    \colhead{(deg)} &
    \colhead{} &
    \colhead{($10^{-9}$)}
  }
  \startdata
  Mercury &    87.97 &  0.39 & 0.206 &  29.1 &  0.38 & 83.0 & 0.11 & 0.172 \\
  Venus   &   224.70 &  0.72 & 0.007 &  55.2 &  0.94 & 86.6 & 0.65 & 2.000 \\
  Earth   &   365.26 &  1.00 & 0.017 & 114.2 &  1.00 & 90.0 & 0.37 & 0.646 \\
  Mars    &   686.98 &  1.52 & 0.093 & 286.5 &  0.53 & 88.2 & 0.15 & 0.040 \\
  Jupiter &  4332.82 &  5.20 & 0.048 & 275.1 & 11.22 & 88.7 & 0.52 & 4.836 \\
  Saturn  & 10755.70 &  9.54 & 0.054 & 336.0 &  9.46 & 87.5 & 0.47 & 0.880 \\
  Uranus  & 30687.15 & 19.19 & 0.047 &  96.5 &  4.01 & 89.2 & 0.51 & 0.037 \\
  Neptune & 60190.03 & 30.07 & 0.009 & 265.6 &  3.89 & 88.2 & 0.41 & 0.013
  \enddata
\end{deluxetable*}

\begin{figure*}
  \begin{center}
    \includegraphics[angle=270,width=15.0cm]{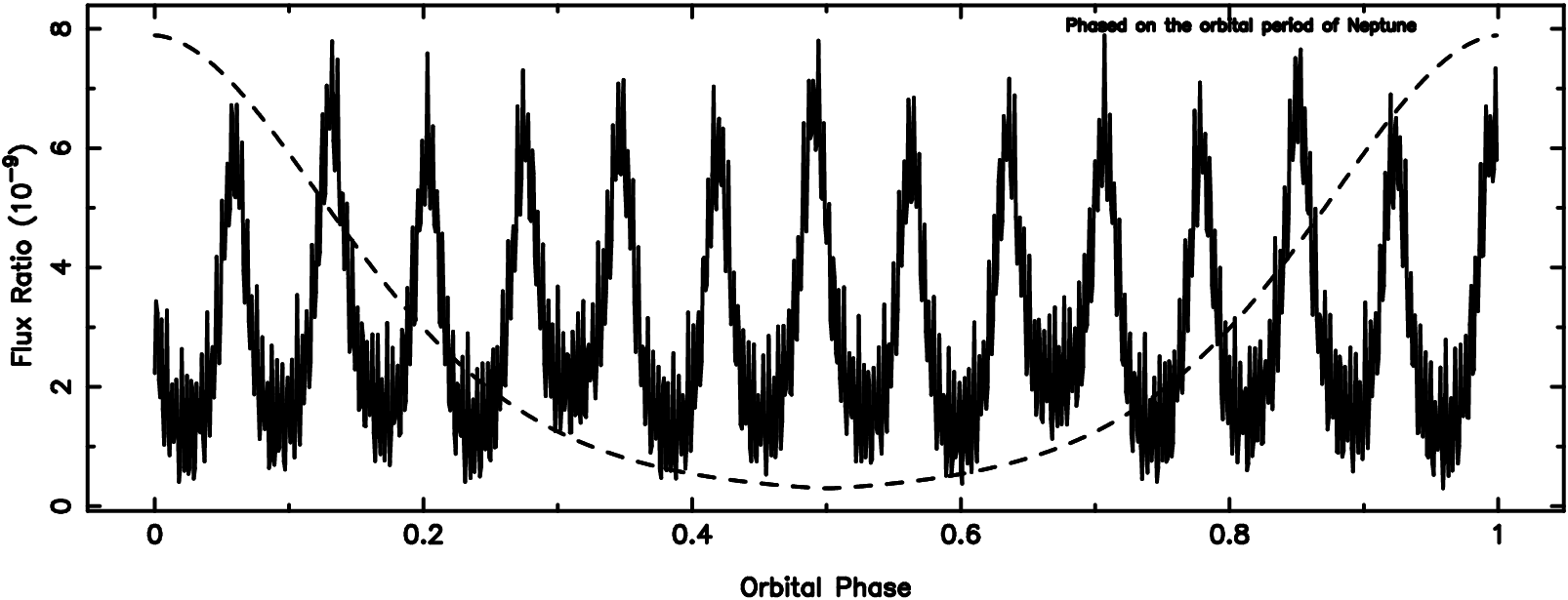} \\
    \includegraphics[angle=270,width=15.0cm]{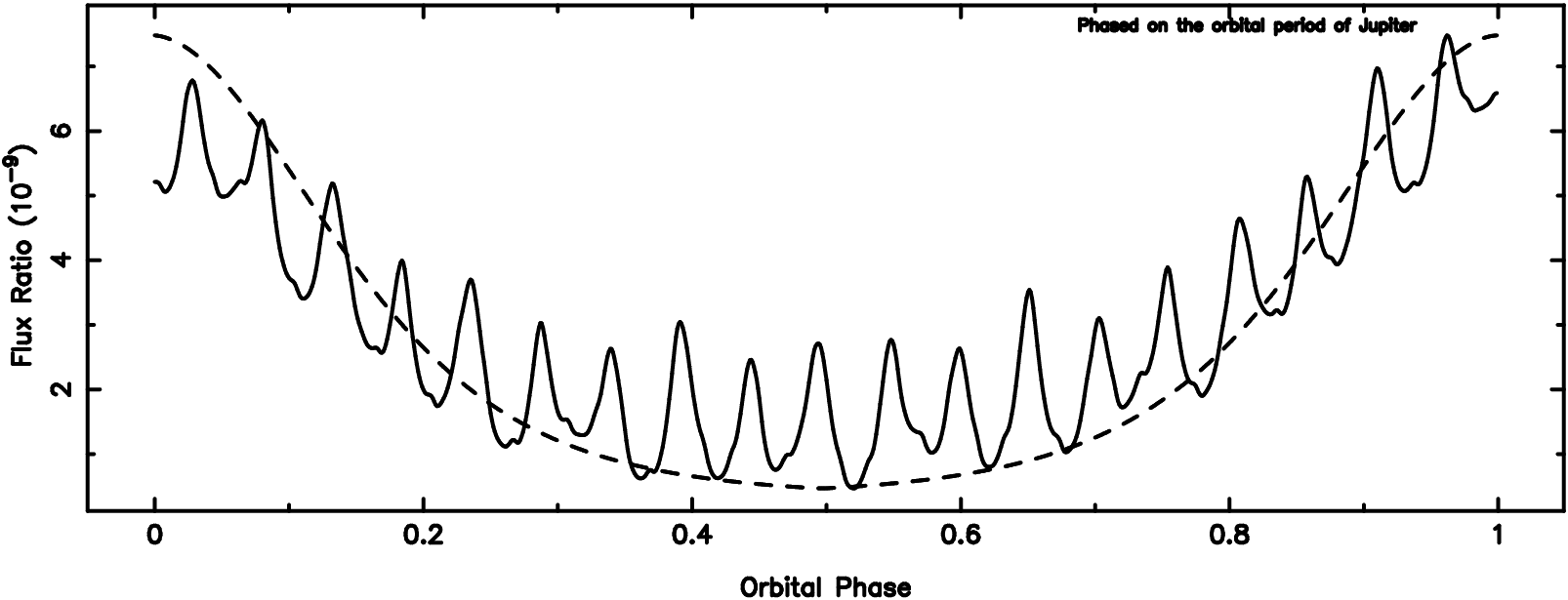}
  \end{center}
  \caption{Predicted photometric flux variation for the Solar System
    as observed along the ecliptic. The top panel is phased on the
    orbital period of Neptune. The bottom panel is phased on the orbit
    of Jupiter to highlight the contributions of the terrestrial
    planets, most notably Venus. The dashed lines in each case
    represent the normalized phase function of Neptune and Jupiter
    respectively.}
  \label{ssphase}
\end{figure*}

Figure \ref{ssphase} shows these phase variations at two resolutions;
one phased on the orbit of Neptune and the other phased on the orbit
of Jupiter. The top panel is dominated by Jupiter which has the
largest phase amplitude. Uranus and Neptune have negligible
contributions to the combined phase variations and Saturn produces a
small modulation in the Jupiter signature. However, the terrestrial
planets are at least as significant in their contributions. Venus has
a particularly large contribution, the amplitude of which is 43\% that
of Jupiter. The effects of the terrestrial planets can be seen much
clearer in the lower panel which shows the phase variations for one
complete orbital period of Jupiter. Reaching the precision to detect
the largest planet in our system is no guarantee that one would see
the effects of the terrestrial planets as anything other than
partially-correlated noise. This is compounded by the relatively short
orbital periods of the inner planets whose relation to the observation
cadence may render those planets all but undetectable. The flux
amplitudes for the Solar System planets are extremely small, but this
is an effect which can be scaled to alternative orbital
configurations. We will see several examples of this in the following
section discussing Kepler systems, where the predicted flux amplitudes
are several orders of magnitude larger.


\subsection{Transit Probabilities}
\label{sstran}

\begin{figure*}
  \begin{center}
    \includegraphics[angle=270,width=15.0cm]{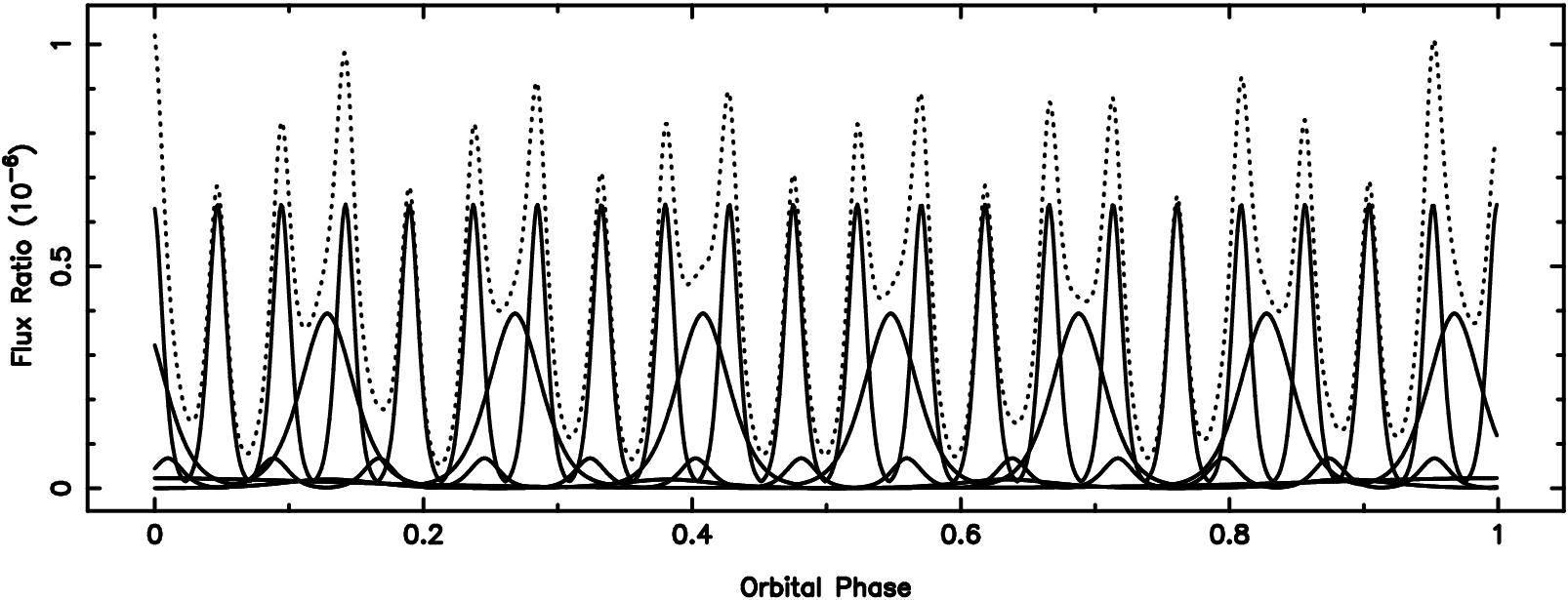} \\
    \includegraphics[angle=270,width=15.0cm]{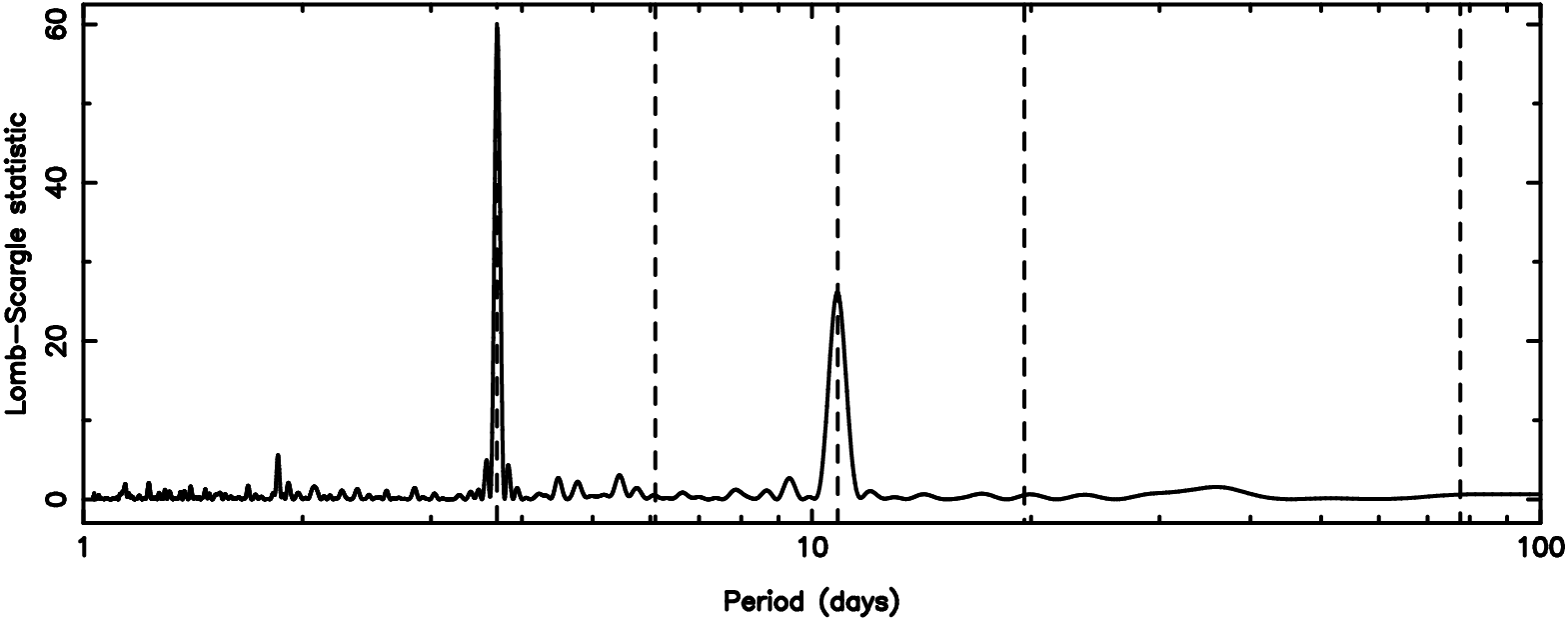}
  \end{center}
  \caption{Top panel: The model photometric flux variations due to
    planetary phases in the Kepler-20 system for one complete orbital
    period of the outer (d) planet. The solid lines show the phase
    variations due to the individual planets and the dotted line
    indicates the combined effect. Bottom panel: The periodogram
    resulting from a fourier analysis of the data. The vertical dashed
    lines indicate the actual periods of the five planets.}
  \label{kep20fig}
\end{figure*}

It is worth noting that not all of the Solar System would be
detectable via the transit method, even if viewed along the ecliptic.
The detection of the outer planets via either the transit or radial
velocity techniques is especially difficult, as described by
\citet{kan11b}. For a circular orbit the geometric transit probability
is inversely proportional to $a$, such that the inclination of the
planet's orbital plane, $i$, must satisfy
\begin{equation}
  a \cos i \leq R_p + R_\star
\end{equation}
By this criteria, observing our Sun along the ecliptic plane would
only result in a transit of Earth. We performed a Monte-Carlo
simulation which rotates the plane of the Solar System $\pm 10\degr$
with respect to the observer. This simulation showed that there is
only a very narrow range of viewing angles for which one would see
more than one planet transit the Sun. Specifically, if the viewing
angle departs from the ecliptic by between 1.76\degr and 1.78\degr,
one will observe transits of both Mars and Neptune. This could be
considered an extreme case of demonstrating that not all planets are
necessarily accounted for in multi-planet transiting systems and thus
the phase variations shown in Figure \ref{ssphase} will not be
representative of what is predicted. However, the issue is still
particularly pertinent for Kepler multi-planet systems which are in
much more compact orbital configurations, as we will see in the
following sections.


\section{Kepler Systems}

Here we apply this methodology to several of the known multi-planet
systems detected by the Kepler mission and discuss the impact on
phase signatures.


\subsection{Kepler-20}

The Kepler-20 system consists of five known transiting planets which
were discovered by \citet{fre12} and \citet{gau12}. Three of the
planets (b, c, and d) are of super-Earth size 1.91--3.07~$R_\oplus$
and the other two (e and f) are Earth size and smaller. The orbital
and physical parameters for these planets are shown in Table
\ref{kep20table}. In the absence of information regarding the
atmospheric properties of these planets, we have no knowledge of their
geometric albedos. The potential diversity of exoplanetary
atmospheres, even amongst those of similar mass and semi-major axis,
is sufficient to disqualify an assumption regarding the albedo of a
planet. However, we are aware of an albedo dependence on star--planet
separation \citep{sud05,cah10,kan10}, such as the case of HD~209458b
which was determined to have an upper limit of $A_g < 0.08$ from
observations using the Microvariability and Oscillations of STars
(MOST) satellite \citep{row08}. For the purposes of this analysis, we
assign a moderately low default value of $A_g = 0.2$ to all planets so
that we may evaluate the relative contributions of the other
parameters to the overall flux. When performing the analysis for a
given dataset, these albedos may be treated as free parameters when
identifying the shape of the phase signatures which match to the
observed orbital periods from the transit photometry.

\begin{deluxetable}{ccccccccc}
  \tablecolumns{9}
  \tablewidth{0pc}
  \tablecaption{\label{kep20table} Kepler-20 Planetary Orbital
    Parameters and Peak Flux Ratios}
  \tablehead{
    \colhead{Planet} &
    \colhead{$P$} &
    \colhead{$a$} &
    \colhead{$R_p$} &
    \colhead{$i$} &
    \colhead{$A_g$} &
    \colhead{Flux Ratio} \\
    \colhead{} &
    \colhead{(days)} &
    \colhead{(AU)} &
    \colhead{($R_\oplus$)} &
    \colhead{(deg)} &
    \colhead{} &
    \colhead{($10^{-6}$)}
  }
  \startdata
  b &  3.696 & 0.045 & 1.91 & 86.50 & 0.2 & 0.639 \\
  c & 10.854 & 0.093 & 3.07 & 88.39 & 0.2 & 0.394 \\
  d & 77.612 & 0.345 & 2.75 & 89.57 & 0.2 & 0.023 \\
  e &  6.098 & 0.063 & 0.87 & 87.50 & 0.2 & 0.068 \\
  f & 19.577 & 0.138 & 1.03 & 88.68 & 0.2 & 0.020
  \enddata
\end{deluxetable}

\begin{figure*}
  \begin{center}
    \begin{tabular}{cc}
      \includegraphics[angle=270,width=8.2cm]{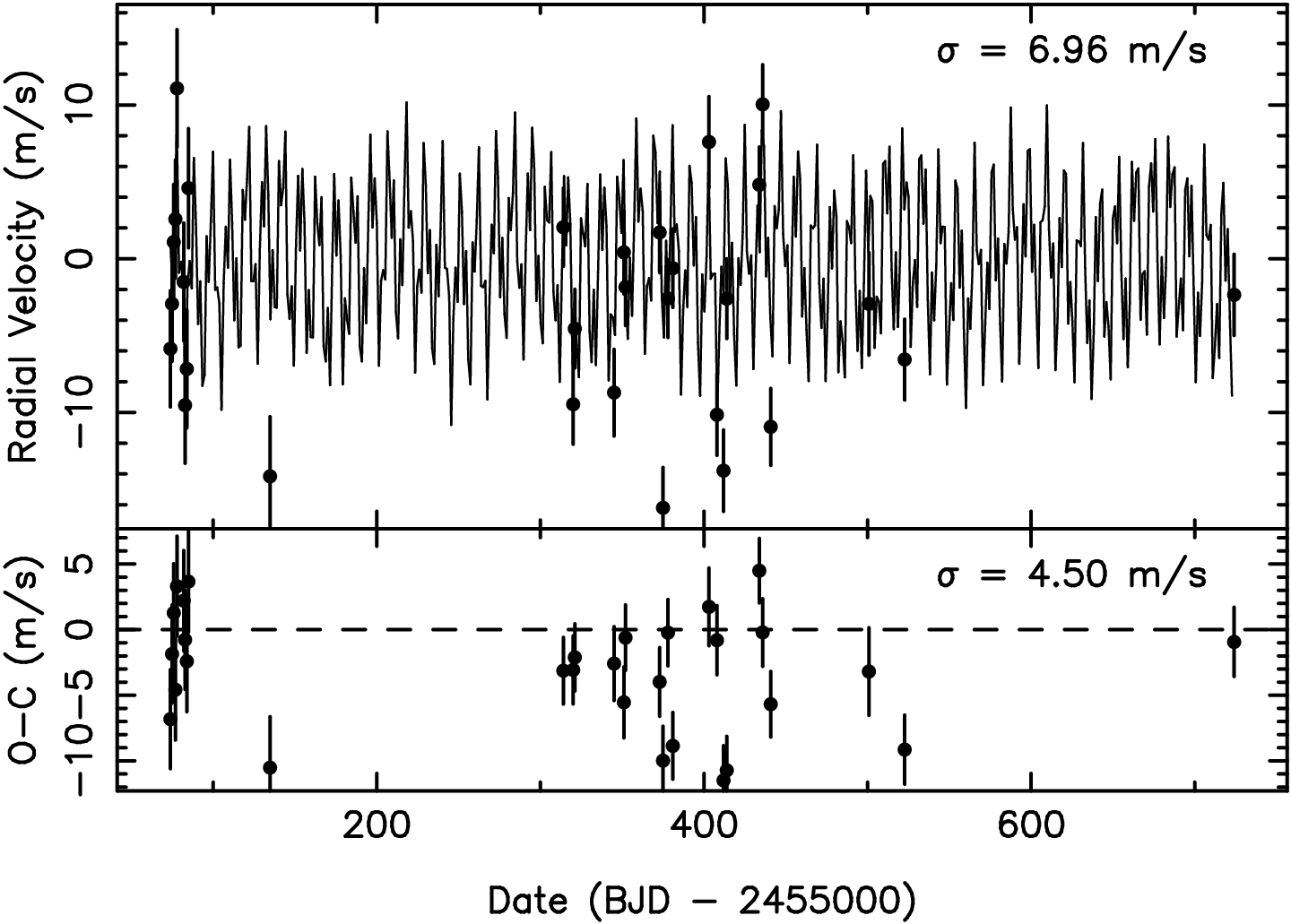} &
      \includegraphics[angle=270,width=8.2cm]{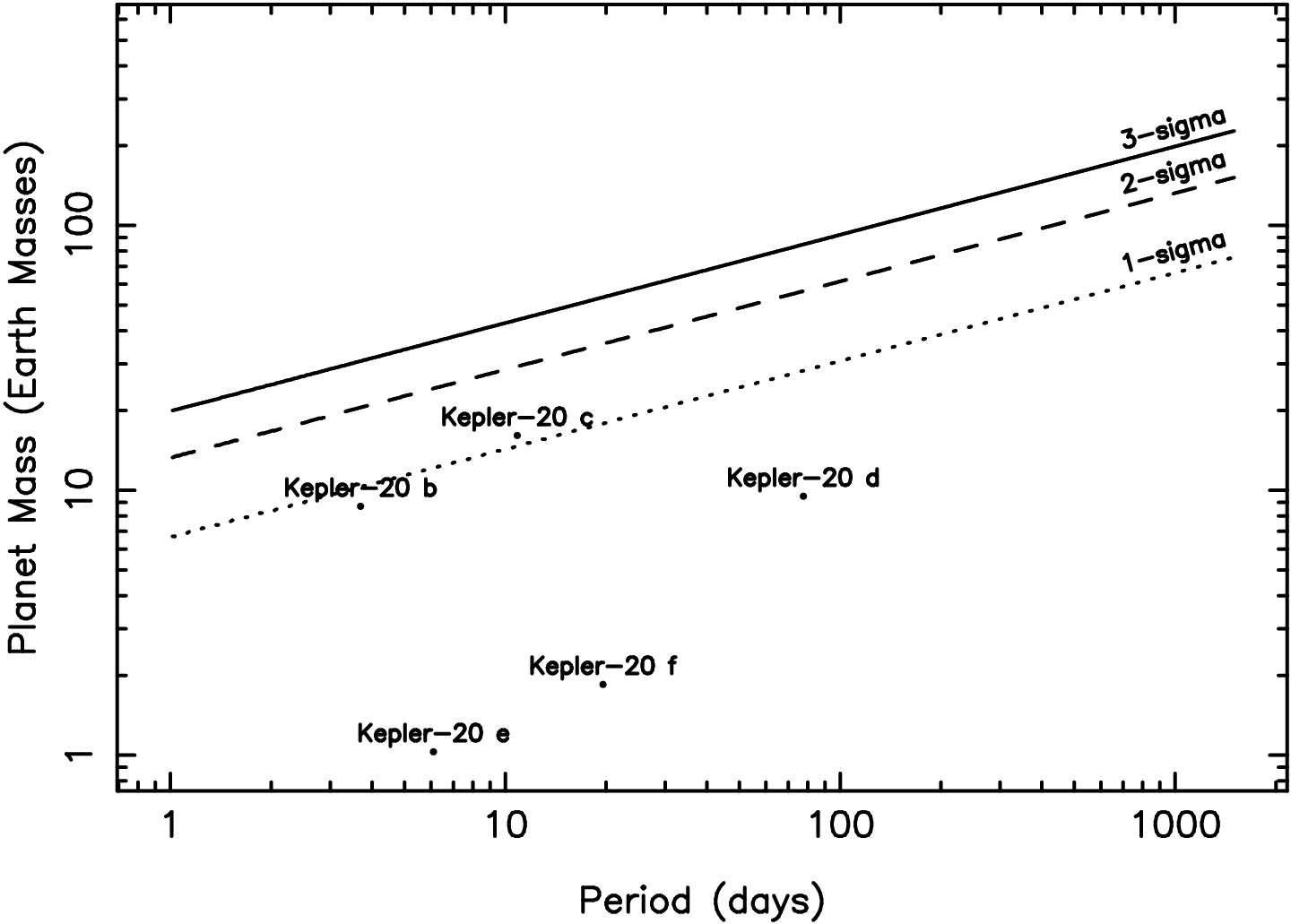}
    \end{tabular}
  \end{center}
  \caption{Left panel: The radial velocity data of the Kepler-20
    system \citep{gau12} with the radial velocity model predicted by
    the planetary parameters. The lower panel shows the residuals
    after subtracting the model from the data. Right panel: Regions of
    exclusion (above the lines) for additional planets within the
    system based upon the rms scatter of the residuals. The detected
    planets in the system are shown for reference.}
  \label{kep20rv}
\end{figure*}

The resulting model phase variations for the Kepler-20 system are
shown in the top panel of Figure \ref{kep20fig}, with the contribution
from individual planets shown as solid lines and the total phase
variations indicated by a dotted line. The two largest contributors
to the phase variation are the b and c planets due to their larger
size and proximity to the star, shown in Table \ref{kep20table}. The e
planet has a smaller star--planet separation than the c planet but the
phase amplitude is inhibited by its small size. However, were the e
planet to have a substantial reflective atmosphere such as Venus, the
amplitude of the phase variation for planet e would rise to become
comparable to around half that of planet c. Thus the two smallest
planets in this system may yet be distinguishable from the major
sources of reflected light in the system.

The bottom panel of Figure \ref{kep20fig} shows a sample periodogram
using the techniques described in Section \ref{fourier}, where we have
simulated data for the model phase curves shown in the top panel. The
vertical dashed lines indicate the orbital periods of the known five
planets in the system. As discussed earlier, in cases such as this
where the phase signature is largely dominated by one or more planets,
the remaining signatures may remain invisible regardless of the
quality of the data or the time baseline of the observations. Thus we
see that the signatures of the b and c planets, at 3.7 and 10.9 days,
are unambiguiously extracted from the data. One possible solution is
to perform a fit to the these two phase signatures, subtracting them
from the data, and then re-analysing the residuals in an attempt to
reveal presence of the remaining planets. The success of such an
approach depends on the amplitude of the remaining phase signatures
with respect to the noise properties of the data.

To investigate the potential for additional sources of phase amplitude
components in the system, we re-analyse the Keck/HIRES radial velocity
data acquired for Kepler-20 by \citet{gau12}. In particular, we would
like to investigate evidence of any additional companions in the
system which may not necessarily transit the host star. Shown in the
left panel of Figure \ref{kep20rv} are the RV data along with the
Keplerian model predicted by the planetary parameters described by
\citet{gau12} and the residuals to the fit shown at the bottom.

The mass of the d planet is difficult to constrain from the RV data
and is described by \citet{gau12} as having a $2\sigma$ limit of $<
20.1 \ M_\oplus$. The Markov Chain Monte Carlo (MCMC) analysis of the
RV data yields a mass estimate for the d planet of
$9.50^{+6.23}_{-8.17} \ M_\oplus$ (Geoff Marcy, private
communication). We adopt this value of the mass for deriving the
Keplerian model for the data which then produces the shown
residuals. The predicted RV semi-amplitudes for the e and f planets
are 0.38~m/s and 0.47~m/s respectively and therefore have a negligible
contribution to the model.

The residuals shown in Figure \ref{kep20rv} still contain a slightly
reduced rms scatter of 4.5~m/s. We calculate RV amplitudes that equal
this $1\sigma$ scatter as a function of orbital period and planet
mass. These are shown in the right panel of Figure \ref{kep20rv}, with
the lines representing multiples of the rms scatter of the
residuals. One consequence of this figure is that we can exclude the
presence of a Jupiter-mass planet in the system out to a period of
$\sim 1000$~days at the $3\sigma$ level. The exclusion yields
significant information for a correct model of the phase variations
since a Jupiter-size planet with a 200 day period would produce a
long-term trend in the photometry, where the amplitude is similar to
that of the c planet.


\subsection{Kepler-33}

\begin{figure*}
  \begin{center}
    \includegraphics[angle=270,width=15.0cm]{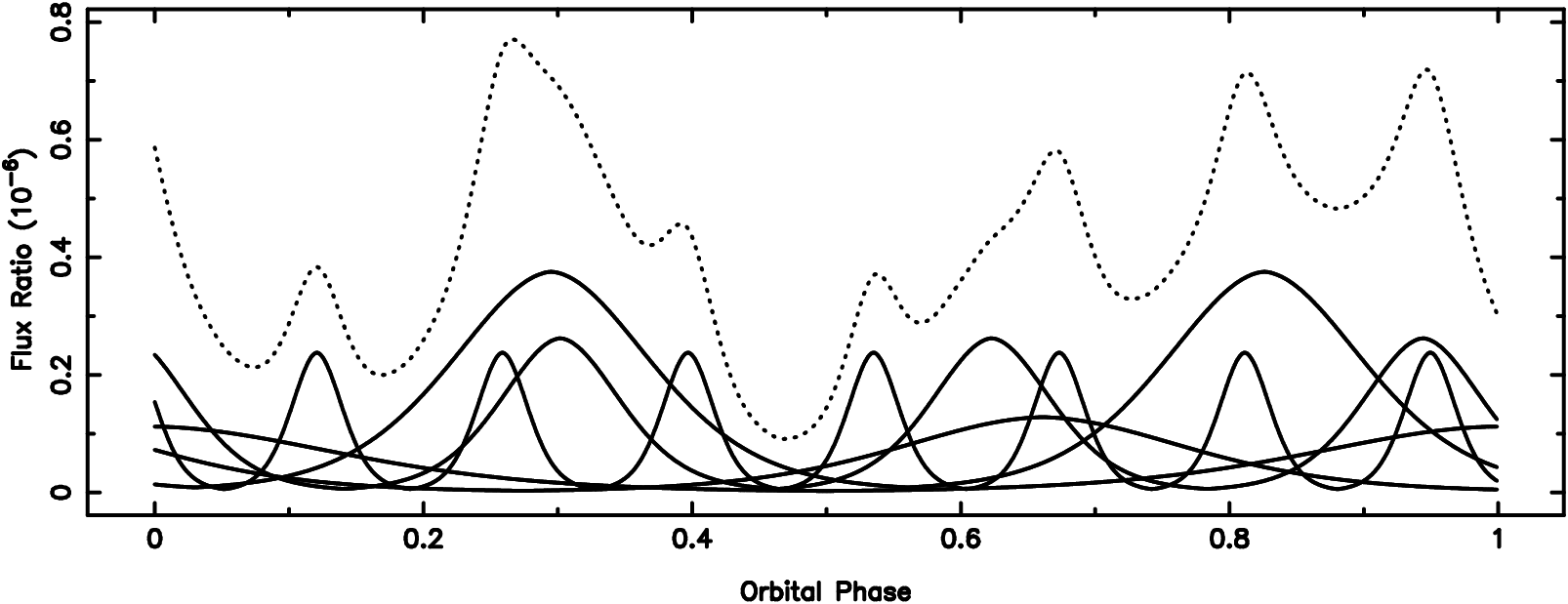} \\
    \includegraphics[angle=270,width=15.0cm]{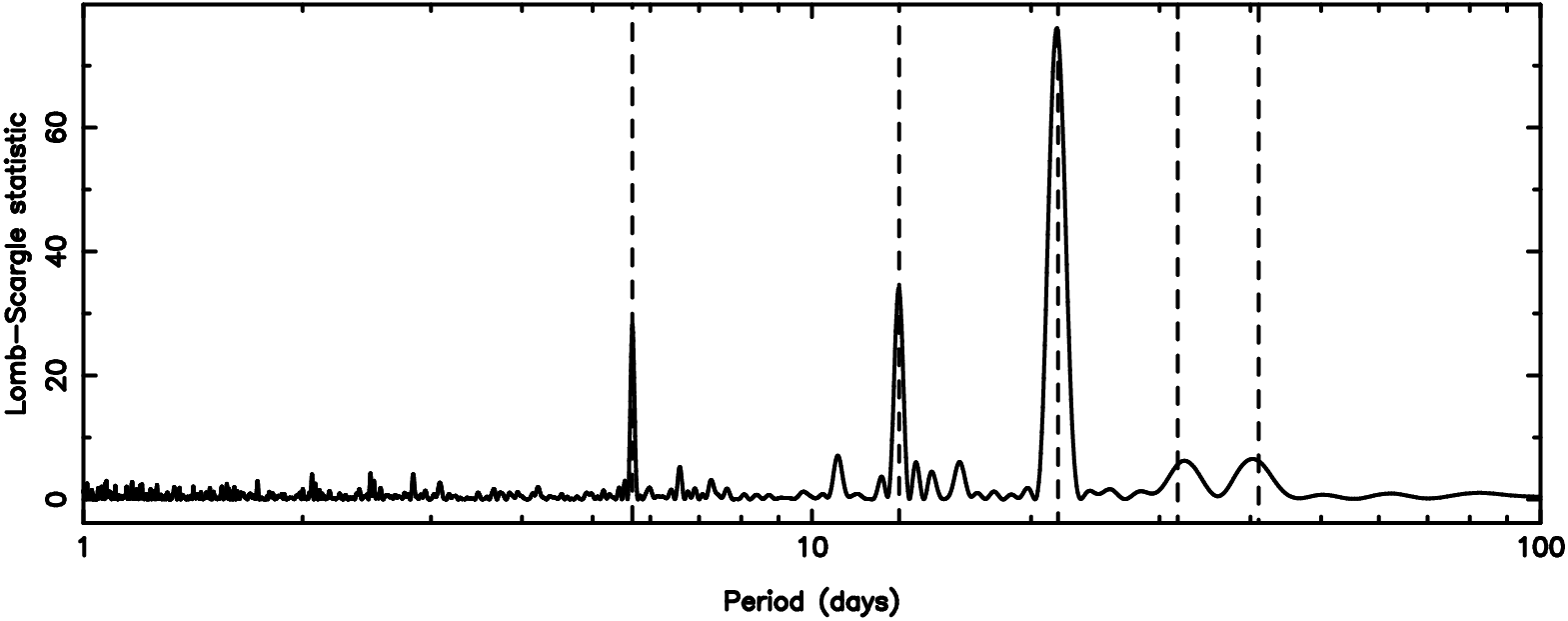}
  \end{center}
  \caption{Top panel: The model photometric flux variations due to
    planetary phases in the Kepler-33 system for one complete orbital
    period of the outer (f) planet. The solid lines show the phase
    variations due to the individual planets and the dotted line
    indicates the combined effect. Bottom panel: The periodogram
    resulting from a fourier analysis of the data. The vertical dashed
    lines indicate the actual periods of the five planets.}
  \label{kep33fig}
\end{figure*}

Kepler-33 is a system of five known transiting planets which was
announced by \citet{lis12}. The complete parameters for these planets
are shown in Table \ref{kep33table}, where we have once again adopted
fiducial albedo values of 0.2. This system is quite different from the
Kepler-20 system since it consists of planets in a size regime ranging
from super-Earth to Neptune and a compact period range between 5 and
42 days. The major consequence of this compact planetary configuration
is that all of the detected planets contribute comparable amounts of
flux to the total phase variations of the system. For example, since
the outer planets have a larger size, it somewhat compensates for
their larger star--planet separation. The phase variations are shown
in the top panel of Figure \ref{kep33fig}, where the solid lines
indicate the individual planets and the dotted line represents the
combined effect. The similar phase amplitudes and small range of
periods results in a significantly improved fourier decoupling of the
phase signatures compared with the Kepler-20 system. The bottom panel
of Figure \ref{kep33fig} shows the periodogram from a simulated
dataset constructed from the phase curves shown in the top panel. The
inner three planets are recovered with ease due to their relatively
large phase amplitudes. The outer two planets have detectable
signatures from the combined analysis of all five planets but would
benefit from the prior extraction of the three dominant planets, as
discussed earlier.

\begin{deluxetable}{ccccccccc}
  \tablecolumns{9}
  \tablewidth{0pc}
  \tablecaption{\label{kep33table} Kepler-33 Planetary Orbital
    Parameters and Peak Flux Ratios}
  \tablehead{
    \colhead{Planet} &
    \colhead{$P$} &
    \colhead{$a$} &
    \colhead{$R_p$} &
    \colhead{$i$} &
    \colhead{$A_g$} &
    \colhead{Flux Ratio} \\
    \colhead{} &
    \colhead{(days)} &
    \colhead{(AU)} &
    \colhead{($R_\oplus$)} &
    \colhead{(deg)} &
    \colhead{} &
    \colhead{($10^{-6}$)}
  }
  \startdata
  b &  5.668 & 0.068 & 1.74 & 86.39 & 0.2 & 0.238 \\
  c & 13.176 & 0.119 & 3.20 & 88.19 & 0.2 & 0.262 \\
  d & 21.776 & 0.166 & 5.35 & 88.71 & 0.2 & 0.375 \\
  e & 31.784 & 0.214 & 4.02 & 88.94 & 0.2 & 0.128 \\
  f & 41.029 & 0.254 & 4.46 & 89.17 & 0.2 & 0.112
  \enddata
\end{deluxetable}

\begin{figure*}
  \begin{center}
    \includegraphics[angle=270,width=15.0cm]{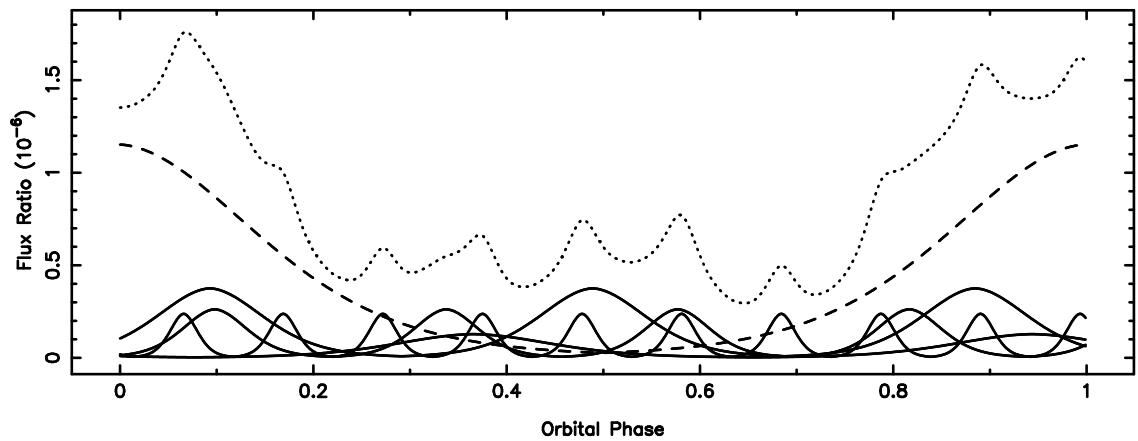}
  \end{center}
  \caption{As for the top panel of Figure \ref{kep33fig} except that
    an additional non-transiting planet has been added with a period
    of 55 days, indicated by the dashed line. The plot now shows one
    complete orbital period of this new planet.}
  \label{kep33planet}
\end{figure*}

RV data is not available for Kepler-33 at the time of writing, which
prevents the exclusion of additional planets as was performed for the
Kepler-20 system. For such a compact system, it is not unreasonable to
postulate the existence of further non-transiting planets in the
system. This could have a profound effect on the combined phase
variations if Jovian-size planets lie just beyond the reach of the
transit method to detect them. An example of such a case is shown in
Figure \ref{kep33planet}. We have added the signature (shown as a
dashed line) of a non-transiting Jupiter-size planet with an orbital
period of 55 days. This planet immediately dominates the phase
variations with an amplitude which is almost a full order of magnitude
larger than the previous largest amplitude of the c planet. The
signatures of the inner planets are thus reduced to correlated noise
in the face of this new dominant signature, making extraction more
difficult since the albedos of all of the planets are treated as free
parameters in the fit to the data.


\section{Conclusions}

Initial results from the Kepler mission indicate that multi-planet
systems are relatively common. The precision of the Kepler mission and
future photometric studies will allow detailed investigations of
photometric variations due to planetary phase signatures. We have
shown that the various phase signatures in these systems can be
combined to produce an emulation of correlated noise in the
photometry. This confusion can both cause the planets to be
indistinguishable from each other and disguise the phase signature
from the dominant planet in the system. On the other hand,
non-transiting planets, usually at star--planet separations beyond the
known size of the system, may either introduce new sources of
correlated noise or indeed be the dominant source of variations in the
system, such as was demonstrated for Kepler-33.

We have shown that fourier analysis of the data will be able to
extract many signatures from multi-planet systems, but depends
critically upon the cadence and duration of the observations in order
for the frequency distribution of the signatures to be properly
sampled. In practice, the largest hinderance will be that of the
photometric precision, which will prevent even high-frequency signals
from being detected. Our investigation of the Solar System
demonstrates an example of confusion sources, where the signals of
Jupiter and Venus may be extracted but the signatures of the remaining
planets are unlikely to be interpreted as anything other than
correlated noise. It is worth noting that the Solar System would not
be detected as a multi-planet system due to the mis-alignment of the
planetary orbital planes, with the exception of Mars and Neptune. We
further demonstrate the difficulties in deciphering the phase
variation for the Kepler-20 and Kepler-33 systems. The interesting
aspect of these systems, along with many other Kepler multi-planet
systems, is the compact nature of the orbits such that it is possible
to observe complete orbital phases for all planets within a reasonable
time frame. One corollary of this is that there is great potential for
a non-transiting giant planet to exist beyond the known planets in
each system whose signature would significantly change the predicted
phase variations of the system.

We have focused solely on the photoemetric variations due to the phase
functions of the planets. There are however numerous other effects,
mentioned in Section \ref{intro}, which need to be accounted for
before one can begin to examine the signals discussed here. With
further observations of the host stars and an extended baseline of
precision photometry, it is hoped that we can eventually disentangle
the various contributions to the observed flux from these most
interesting systems.


\section*{Acknowledgements}

The authors would like to thank Geoff Marcy, Nick Gautier, Natalie
Hinkel, and Diana Dragomir for their useful suggestions and comments.
We would also like to thank the anonymous referee, whose comments
greatly improved the quality of the paper.



\begin{thebibliography}{}

\bibitem[\protect\citeauthoryear{Aharmin et al.}{2005}]{aha05}
  Aharmin, B., et al. 2005, Phys. Rev. D, 72, 052010
\bibitem[\protect\citeauthoryear{Batalha et al.}{2012}]{bat12}
  Batalha, N.M., et al. 2012, ApJS, submitted (arXiv:1202.5852)
\bibitem[\protect\citeauthoryear{Borucki et al.}{2011a}]{bor11a}
  Borucki, W.J., et al. 2011, ApJ, 728, 117
\bibitem[\protect\citeauthoryear{Borucki et al.}{2011b}]{bor11b}
  Borucki, W.J., et al. 2011, ApJ, 736, 19
\bibitem[\protect\citeauthoryear{Cahoy et al.}{2010}]{cah10} Cahoy,
  K.L., Marley, M.S., Fortney, J.J. 2010, ApJ, 724, 189
\bibitem[\protect\citeauthoryear{Cumming}{2004}]{cum04} Cumming,
  A. 2004, MNRAS, 354, 1165
\bibitem[\protect\citeauthoryear{Cumming et al.}{2008}]{cum08}
  Cumming, A., Butler, R.P., Marcy, G.W., Vogt, S.S., Wright, J.T.,
  Fischer, D.A. 2008, PASP, 120, 531
\bibitem[\protect\citeauthoryear{Endl et al.}{2000}]{end00} Endl, M.,
  K\"urster, M., Els, S. 2000, A\&A, 362, 585
\bibitem[\protect\citeauthoryear{Fressin et al.}{2012}]{fre12}
  Fressin, F., et al. 2012, Nature, 482, 195
\bibitem[\protect\citeauthoryear{Gautier et al.}{2012}]{gau12}
  Gautier, T.N., et al. 2012, ApJ, 749, 15
\bibitem[\protect\citeauthoryear{Jackson et al.}{2012}]{jac12}
  Jackson, B.K., Lewis, N.K., Barnes, J.W., Deming D.L., Showman,
  A.P., Fortney, J.J. 2012, ApJ, 751, 112
\bibitem[\protect\citeauthoryear{Kane \& Gelino}{2010}]{kan10} Kane,
  S.R., Gelino, D.M. 2010, ApJ, 724, 818
\bibitem[\protect\citeauthoryear{Kane \& Gelino}{2011}]{kan11a} Kane,
  S.R., Gelino, D.M. 2011, ApJ, 729, 74
\bibitem[\protect\citeauthoryear{Kane}{2011}]{kan11b} Kane, S.R. 2011,
  Icarus, 214, 327
\bibitem[\protect\citeauthoryear{Kane \& Gelino}{2012}]{kan12} Kane,
  S.R., Gelino, D.M. 2012, MNRAS, 424, 779
\bibitem[\protect\citeauthoryear{Lissauer et al.}{2011a}]{lis11a}
  Lissauer, J.J., et al. 2011, Nature, 470, 53
\bibitem[\protect\citeauthoryear{Lissauer et al.}{2011b}]{lis11b}
  Lissauer, J.J., et al. 2011, ApJS, 197, 8
\bibitem[\protect\citeauthoryear{Lissauer et al.}{2012}]{lis12}
  Lissauer, J.J., et al. 2012, ApJ, 750, 112
\bibitem[\protect\citeauthoryear{Lomb}{1976}]{lom76} Lomb, N.R. 1976,
  Ap\&SS, 39, 447
\bibitem[\protect\citeauthoryear{Mazeh et al.}{2012}]{maz12} Mazeh,
  T., Nachmani, G., Sokol, G., Faigler, S., Zucker, S. 2012, A\&A,
  541, 56
\bibitem[\protect\citeauthoryear{Moorhead et al.}{2011}]{mor11}
  Moorhead, A.V., et al. 2011, ApJS, 197, 1
\bibitem[\protect\citeauthoryear{O'Toole et al.}{2009}]{oto09}
  O'Toole, S.J.; Tinney, C.G., Jones, H.R.A., Butler, R.P., Marcy,
  G.W., Carter, B., Bailey, J. 2009, MNRAS, 392, 641
\bibitem[\protect\citeauthoryear{Rowe et al.}{2008}]{row08} Rowe,
  J.F., et al. 2008, ApJ, 689, 1345
\bibitem[\protect\citeauthoryear{Pfahl et al.}{2008}]{pfa08} Pfahl,
  E., Arras, P., Paxton, B. 2008, ApJ, 679, 783
\bibitem[\protect\citeauthoryear{Scargle}{1982}]{sca82} Scargle, J.D.
  1982, ApJ, 263, 835
\bibitem[\protect\citeauthoryear{Shporer et al.}{2011}]{shp11}
  Shporer, A., et al. 2011, AJ, 142, 195
\bibitem[\protect\citeauthoryear{Sudarsky et al.}{2005}]{sud05}
  Sudarsky, D., Burrows, A., Hubeny, I., Li, A. 2005, ApJ, 627, 520
\bibitem[\protect\citeauthoryear{Welsh et al.}{2010}]{wel10} Welsh,
  W.F., Orosz, J.A., Seager, S., Fortney, J.J., Jenkins, J., Rowe,
  J.F., Koch, D., Borucki, W.J. 2010, ApJ, 713, L145
\bibitem[\protect\citeauthoryear{Wright et al.}{2009}]{wri09} Wright,
  J.T., Upadhyay, S., Marcy, G.W., Fischer, D.A., Ford, E.B., Johnson,
  J.A. 2009, ApJ, 693, 1084

\end{thebibliography}
\end{document}